\documentstyle[preprint,eqsecnum,aps,epsf,psfig]{revtex}
\newif\iftightenlines\tightenlinesfalse
\tightenlines\tightenlinestrue
\begin{document}
\draft
\title{Charged heavy vector boson production
at the Large Hadron Collider}
\author{Dal Soo Oh and M. H. Reno}
\address{
Department of Physics and Astronomy, University of Iowa, Iowa City,
Iowa 52242}

\maketitle

\begin{abstract}
We evaluate the sensitivity of the Large Hadron Collider (LHC)
to charged
heavy vector boson production followed by their decays to 
$W^\pm Z^0$. We include the
correlated decays of the gauge bosons to leptonic final states.
With an integrated luminosity of $10^5$ pb$^{-1}$, charged technirhos
in the minimal $SU(N)_{TC}$ models for $N\geq 7$ yield signals
with a significance larger than 5. In more general models, we explore
the range of parameter space to which LHC experiments will be sensitive.
Rapidity correlations exhibiting enhanced longitudinal gauge bson
pair production are also shown.

\end{abstract}

\section{INTRODUCTION}

Anomalous weak diboson production at 
lepton\cite{lepton} and hadron\cite{hadron,wznlo}
colliders has been a topic of considerable
interest because of its potential
for exhibiting physics beyond the standard model at the
TeV scale. In particular, non-standard physics may modify tri-boson
couplings, yielding enhanced production of two electroweak gauge bosons.
Modifications of the tri-boson couplings change the balance of purely
longitudinally polarized final state bosons, purely transverse, and
mixed polarization final states.
Extracting vector boson polarizations at the Large Hadron Collider
(LHC) will be
a challenge. Our purpose here is to explore the enhancement of
$W_L Z_L$ via their lepton decay modes
in models with heavy vector boson intermediate states with couplings to
longitudinal gauge bosons at the LHC.

Longitudinal weak boson enhancement is accomplished here with
the introduction of a spin-one, isospin-one resonance ($V$) such as
the technirho ($\rho_{T}$) that occurs in technicolor models \cite{dimop}.
Technicolor
offers an alternative model
to the standard model high energy behavior
which is dominated by transverse gauge boson production. The enhancement
of the cross section of longitudinal weak boson pairs
comes because of $s$-channel production of technirho particles.
We look at the  
charged technirho $\rho_T^\pm\rightarrow W^\pm Z$
and its generalization to $V^\pm\rightarrow W^\pm Z$
\cite{bess}. The advantage of this
channel over the neutral channel: $V^0\rightarrow W^+ W^-$ is that
the neutral channel is plagued by the QCD background coming from
$pp\rightarrow t\bar{t}\rightarrow W^+W^-b\bar{b}$.

Compelling technicolor models which can account for mass generation in
the fermion as well as the gauge boson sector are difficult to construct.
We look at generic features of the coupling of a spin-one isospin-one
vector boson which in a certain limit corresponds to
the minimal technicolor model
with one SU(2) doublet of technifermions that transforms in the fundamental
representation of the SU($N$)$_{TC}$ gauge group \cite{dimop}.
In all of our calculations, we assume that the direct coupling of $V$ to
fermions is negligible. Direct couplings to fermions would only improve our
calculated signal rates.

Early analyses of vector boson production at high energy hadron colliders
were done by
Eichten, Hinchliffe, Lane and Quigg \cite{ehlq}, Chanowitz and Gaillard 
\cite{cg} and Chivukula \cite{sekhar}.
Later work 
on hadron collider production of $V^\pm$ includes 
Refs. \cite{bagger,trho,casalbuoni,dobado,chanowitz}.
These analyses focused on vector boson distributions or a limited range
of parameter space, including decays to leptons.
Here, we include correlated gauge
boson decays to leptons. We focus on
the extent to which enhanced longitudinal boson production
can be detected for a range of parameters which include the
parameters of the minimal technicolor model. In addition to evaluating
transverse mass distributions, we also exhibit rapidity correlations.

We find that the LHC has a limited range in parameter space for which
heavy vector boson production will be clearly distinguishable from the
standard model. Already by masses of 1.4 TeV, the minimal technicolor
model is inaccessible with $10^5$ pb$^{-1}$ of integrated luminosity.
We show that rapidity correlations will be a useful tool auxiliary
to transverse mass distributions in the exploration of enhanced LL
gauge boson pair production.
In the next section we outline the parameters of the heavy vector
boson models and the changes to the matrix elements required.
In Section III we describe our results, followed by our conclusions
in Section IV.

\section{Di-boson production}

A strongly-interacting symmetry breaking sector can be described in
terms of an effective Lagrangian. The $V$ particles are not fundamental 
particles, but composites of strongly interacting fermions. They should mix 
with photons, $W$'s and $Z$'s. Since we are focusing only on
$V^\pm$, the mixing is with $W^\pm$. A generalization of minimal
technicolor involving the heavy vector boson sector is described
by the BESS (Breaking Electroweak Symmetry Strongly) model \cite{bess}.
In this model, the $V$ boson mass is
\begin{equation}
M_V^2={1\over 4} v^2 a g''^2
\end{equation}
in terms of $v=246$ GeV, a new coupling constant $g''$ and an
arbitrary parameter $a$. Its width, in the $M_V\gg M_W,\ M_Z$ limit,
is
\begin{equation}
\Gamma_V={a M_V^3\over 192\pi v^2}\ .
\end{equation}

Technicolor models follow from an effective Lagrangian which is
assumed to scale directly from QCD dynamics
involving the $\rho$. The Kawarabayshi-Suzuki-Fayyazudin-Riazuddin
(KSFR) relations \cite{ksfr} between the mass and width of the vector
boson
should hold. They are satisfied when $a=2$. Furthermore, the mass of
the technirho is related to the $N$ in $SU(N)_{TC}$ technicolor symmetry
via
\begin{equation}
M_{\rho_T}=M_\rho{F_{\pi_T}\over F_\pi}\Biggl[{3\over N}\Biggr]^{1/2}
\simeq 2\ {\rm TeV}\ \Biggl[{3\over N}\Biggr]^{1/2}
\end{equation}
in the minimal one technifermion $SU(2)$ doublet version assumed here.
In Eq. (2.3),
$F_{\pi_T}=v$, $F_\pi=93$ MeV is the pion decay constant and $M_\rho$ is the
ordinary $\rho$ mass. The decay width of $\rho_T^\pm$ to
the would-be goldstone boson pairs, namely, approximately the
longitudinal boson pairs $W_L^\pm Z_L^0$, is
$
\Gamma(\rho_T^\pm\rightarrow W_L^\pm Z_L^0)$, which
scales as $
({3/ N})^{3/2}$
in the large $M_\rho$ limit.
In these models, $\Gamma(\rho_T^\pm\rightarrow W_L^\pm Z_L^0)$ is the
total decay width $\Gamma_{\rho_T^\pm}$. In the BESS model, the
heavy vector boson decays only to $W_L^\pm Z_L^0$.

In our calculations below, we display results for
minimal technicolor with several
representative values of
$M_{\rho_T^\pm}$ (several values of $N$).
We evaluate technirho
production for $N=3-8$, corresponding to masses
$M_{\rho^\pm}=1.25-2.04$ TeV.
In addition to considering
$\Gamma(\rho_T^\pm\rightarrow W_L^\pm Z_L^0)$ given by Eq. (2.2)
with $a=2$,
we consider a range of values  for
$a$ in Eq. (2.2).
The mass range evaluated here for the BESS generalization
of technicolor ($a\neq 2$) is from 1--2.04 TeV.
We generically represent the heavy vector resonance by $V^{\pm}$.
We now turn to the matrix element squared.

As a consequence of the strongly interacting sector, the $V^\pm$
can mix with $W^\pm$, and in $s$-channel interactions,
enhance longitudinal gauge boson production as the parton
invariant mass squared $s$ approaches $M_V^2$.
In the process considered here  ($pp\rightarrow WZ$), quarks, antiquarks
and intermediate vector bosons are considered partons. 
Production of $WZ$ can occur via $q\bar{q}'$ annihilation \cite{bsm}
and by
vector boson scattering. We restrict
our analysis to leading-order in QCD.
Baur, Han and Ohnemus \cite{wznlo} have shown that the next-to-leading
order standard model distributions for $WZ$ production 
for quark-antiquark annihilation closely follow
the leading order results in the $W^\pm Z^0+0$ jet case, where a jet
requires $p_T(j)>50$ GeV and $\eta (j)<3$ \cite{fnr}.
For the subprocess $WZ\rightarrow WZ$ scattering, 
we use the effective-vector-boson
approximation \cite{evba,kuss} (EVBA).
Consequently, forward jet tagging, as discussed in 
for example Ref. \cite{bagger},
cannot be implemented. As a practical matter, the quark-antiquark
annihilation process dominates.

For $q\bar{q}'$ annihilation, we have
calculated the polarization cross sections $\sigma(q\bar{q}'\rightarrow
W_i Z_j)$ for
$i,j=T,L$ (transverse and longitudinal) polarizations directly
using Form, \cite{vermaseren}  and using the program MadGraph \cite{stelzer}.
The charged heavy vector boson modification of the cross section amounts to a
substitution of the standard model matrix element squared for
$q_i\bar{q}_j\rightarrow W^+_L Z_L$ by
\begin{equation}
|M(q_i\bar{q}_j\rightarrow W^\pm_L Z^0_L)|^2\rightarrow
|M(q_i\bar{q}_j\rightarrow W^\pm_L Z^0_L)|^2 \times\, {a^2\over 4}\,
 {M_V^4\over (\hat{s}-M_V^2)^2+ M_V^2 \Gamma_V^2} 
\end{equation}
in terms of the mass $M_V$, width $\Gamma_V$, parameter $a$
and parton center of mass energy squared $\hat{s}$. This is
the conventional vector meson dominance 
(VDM) form accounting for $W^\pm -
V^\pm$ mixing.

Vector boson scattering occurs in the standard model, however, it is not
as large as $q\bar{q}'$ production of $WZ$ at the LHC. 
The fusion of $W^\pm \gamma$ and $W^\pm Z^0$ intermediate states create
$W^\pm Z^0$ final states. Using the improved vector boson approximation
\cite{kuss} which is a good representation of the perturbative process
$q_i\bar{q}_i'\rightarrow q_f\bar{q}_f'(W^\pm \gamma,W^\pm Z^0 )\rightarrow
q_f\bar{q}_f' W^\pm Z^0$ \cite{bagger}, Kuss and Nuss \cite{kn} have shown
that the vector boson scattering contribution to $\sigma (
pp\rightarrow W^\pm Z^0)$ between $WZ $ invariant masses of 0.5 TeV-2.0 TeV
and pseudo-rapidity cuts of 2.5 is between 10-20\% of the quark-antiquark
annihilation contribution. The EVBA in the leading log approximation
does well for the longitudinal scattering \cite{kn}.
Consequently, we only consider
vector boson scattering contributions involving longitudinal vector
bosons in both the initial and final states in the case of enhanced
couplings in the leading log approximation of the
EVBA. 

Using a generalized model of
chiral coupling to vector mesons, the longitudinal matrix
element is \cite{trho}
\begin{equation}
M(W^\pm_LZ^0_L \rightarrow W^\pm_LZ^0_L)=\Bigl( 1-{3\over 4} a\Bigr)
\Biggl({\hat{t}\over v^2}\Biggr)+{a\over 4}{M_V^2\over v^2}\Biggl(
{\hat{u}-\hat{t}\over \hat{s}-M_V^2+iM_V\Gamma_V} + {\hat{s}-\hat{t}
\over \hat{u}-M_V^2}
\Biggr)\ ,
\end{equation}
in terms of Mandelstam variables $\hat{s},\ \hat{t}$ and $\hat{u}$, 
which are combinations
of $W$ and $Z$ momenta.
The longitudinal matrix element potentially has unitarity problems
associated with the behavior at large $\hat{s}$. In particular, the
isospin-two, spin-zero partial wave amplitude has no resonant structure.
Following Refs. \cite{bagger,trho}, by setting the real part of the two
body partial wave amplitude to less than 1/2 for energies up to $\sqrt{
\hat{s}}=1.5 M_V$, the parameter $a$ is constrained to be in the
range 
\begin{equation}
{\rm Max}(3-4.1\Biggl( {{\rm TeV}\over M_V}\Biggr)^2,\ 0)\leq a\leq
3+4.1\Biggl( {{\rm TeV}\over M_V}\Biggr)^2\ .
\end{equation}
We demonstrate below that the LHC will be sensitive to a range of $a$ 
consistent with the unitarity limit.

The parton differential cross section in
the parton center of mass frame is
\begin{equation}
{d\hat{\sigma}\over d\cos\theta} = {1\over 32\pi\, \hat{s}}|M|^2\ .
\end{equation}
The hadron-hadron cross section is 
\begin{equation}
\sigma = \sum_{ij}\int dx_1\, dx_2\, d\cos\theta f_{i/p} (x_1,Q^2)
f_{j/p}(x_2,Q^2) {d\hat{\sigma}\over d\cos\theta}
\end{equation}
where $i,j$ include sums over quark flavors and the $W_LZ_L$ initial
states. We take $Q=\sqrt{\hat{s}}$
in the MRSA parton distribution functions\cite{mrsa}.
The calculations are done at $\sqrt{S}=14$ TeV in proton-proton collisions.

The decay modes of interest are purely leptonic. We count a single charged
lepton decay mode for each boson decay. In dealing with $Z\rightarrow
l^+l^-$, we assume that the $Z$ four-momentum is well reconstructed.
Since we have evaluated each polarization matrix element separately
for $pp\rightarrow WZ$, we include the polarized $W\rightarrow l \nu$
decay matrix elements. In all cases, we sum over both charges of the
$W$ boson.

\section{Results}

We begin by showing in Figs. 1 and 2 the transverse mass distributions
including technirho production ($a=2$)
for $pp\rightarrow W^\pm Z^0\rightarrow
l_1\nu_1 l_2\bar{l}_2$, assuming $N=8$ (Fig. 1) and $N=3$
(Fig. 2). Only one channel for the $W$ boson decay and the $Z$ boson decay
is included.
The transverse mass is defined by
\begin{equation}
m_T^2= [(\vec{p}_T^{\,2}(Z)+m_Z^2)^{1/2}+((\vec{p}_T(l_1)+\vec{p\llap/}_T)^2
+m_W^2)^{1/2}]^2-(\vec{p}_T(Z)+\vec{p}_T(l_1)+\vec{p\llap/}_T)^2\ .
\end{equation}
The neutrino transverse momentum is indicated by $\vec{p\llap/}_T$.
The dominant background source to heavy boson production followed by decay
to $W^\pm Z^0$
is the transverse boson pair production.
The background
is reduced by requiring the $Z$ and the charged lepton from
$W$ decay to lie in the central region of the detector.
In all of the results shown, we have imposed the following rapidity
cuts:
\begin{eqnarray}
|y_Z| &<& 2.5\ ,\\ \nonumber
|y_{l_1}| & < & 2.5\ ,
\end{eqnarray}
where the lepton $l_1$ comes from $W^\pm$ decay.

Transverse mass plots have the separate contributions by
transverse-transverse (TT) $WZ$ bosons indicated by
the solid line,
longitudinal-longitudinal (LL) by dashed lines
and mixed (TL+LT) contributions shown by dot-dashed lines.
The polarizations are defined in the parton center-of-mass frame.
The contribution labeled LL-VDM comes from $q\bar{q}'$ annihilation to
a heavy technirho, while the LL-VBS contribution comes from
vector boson scattering as in Eq. (2.5). The dashed line labeled
LL is the standard model LL result. 
The heavy solid line is the sum of the TT, TL+LT and nonstandard
LL contributions.

As can be seen from Fig. 1, low mass ($N=8,\ M_\rho=1.25$ TeV) technirho
production will be visible as a peak in the transverse mass distribution.
Vector boson scattering has a small contribution at $\sqrt{S}=14$ TeV. 
For $M_\rho=2.04$ TeV
($N=3$), the transverse mass peak is less well defined, as the
technirho width is 466 GeV in the minimal technicolor model. Statistics
will play an important role for the higher mass range.

In Table I, we show the cross section for $pp\rightarrow 
W^\pm Z^0\rightarrow
l_1\nu_1 l_2\bar{l}_2$
at the LHC for $m_T\geq 0.5$ TeV, in units of $10^{-4}$ pb, 
including technicolor $N=8$ (TC8), $N=4$ (TC4) and $N=3$ (TC3)
for representative nonstandard model values. Both
quark-antiquark annihilation and vector boson scattering are included
in the nonstandard model results. For the proposed integrated luminosity
of the LHC, $\int dt {\cal L}=10^5$ pb$^{-1}$, the number of 
events that come from diagrams with a technirho
is on the order of $\sim 20-80$ events, depending on the value
of the mass. The standard model prediction for the same integrated luminosity
and minimum transverse mass is $\sim 70$ events.

Table II shows the minimum luminosity required to detect
the technirho with a significance,
\begin{equation}
{\rm Significance}={S\over \sqrt{S+B}}\geq 5.
\end{equation}
We define the signal to be the difference between the LL distributions
with and without nonstandard model modifications. The background is the sum
of all standard model contributions to the distribution. 
We have chosen transverse mass cuts to maximize the significance,
limited by our bin size of 25 GeV.
The minimum transverse mass which maximizes the significance varies from 
725 GeV ($N=8$) to 1075 GeV ($N=3$).
The $N=7$
and $N=8$ technicolor models are accessible to the LHC in this
single channel final state, given an integrated luminosity
$\int dt {\cal L}=10^5$ pb$^{-1}$ and the requirement that the
significance is greater than 5.

An interesting distinction between standard model and technirho enhanced
LL production is seen in 
$WZ$ rapidity correlations. Baur, Han and Ohnemus
\cite{zeros} have shown that there are zeros in 
leading order matrix elements
contributing to the TT cross section. Nonstandard LL enhancement fills in
the zeros.
Next-to-leading order contributions
also tend to fill in the zeros, but by imposing a zero-jet cut, the NLO
effects can be largely ignored \cite{wznlo}, as mentioned above. 
Since the $W$ decays to a charged lepton and neutrino, some of the
correlation is washed out, nevertheless, there is a pronounced
difference in the shape of the distribution in
\begin{equation}
\Delta y_{Zl} \equiv y_Z-y_{l_1}\ .
\end{equation}
Figs. 3 and 4 show the cross section $\Delta\sigma$ in each bin in
$\Delta y_{Zl}$ of size 0.2 for $N=8$ and $N=3$, respectively.
The solid line shows the standard model result, while the dashed line
shows the sum of the TT, TL+LT and nonstandard LL. The rapidity cuts
in Eq. (3.2) are imposed, and the transverse mass is restricted to
yield the highest significance. 
For Fig. 3, 725 GeV$\leq m_T\leq 1600$ GeV, while for Fig. 4, 
1075 GeV$\leq m_T\leq 2400$ GeV.

The shapes of the $\Delta y_{Zl}$ distributions for the standard model
and technirho are dramatically different. In Fig. 3, the nonstandard
curve is peaked at $\Delta y_{Zl}=0$ and changes by a factor
of $\sim 5$ in the range from $|\Delta y_{Zl}|=2$ to 
$|\Delta y_{Zl}|=0$. The standard curve changes by less than a factor
of 1.25 and exhibits a slight dip at $\Delta y_{Zl}=0$. The corresponding
change in Fig. 4 for the non-standard model is more than a factor of 3
while the standard model changes by less than a factor of 1.5.

In the more general context of the BESS model, 
the mass and width are related by the parameter $a$. In addition, with
increasing $a$, heavy vector boson production is enhanced. For large
enough $a$, even a heavy vector boson of mass 2 TeV is detectable at the
LHC with $10^5$ pb$^{-1}$ of integrated luminosity.
In Table III we show the minimum $a$ ($a_{min}$)
that gives a significance
$S/\sqrt{S+B}\geq 5$. We have applied the rapidity cuts from Eq. (3.2).
The transverse mass cuts are fixed to maximize the significance
of the signal, however, not set lower than 500 GeV. The resulting values
of $a_{min}$ range from 0.42-3.9 for $m_V=1.00-2.04$ TeV.
These values of $a_{min}$ lie within the limits required by partial wave
unitarity in Eq. (2.6).

Figs. 5 and 6 show the transverse mass distributions for $m_V=1.00$ TeV,
$a=0.42$
and $m_V=2.04$ TeV, $a=3.9$, respectively. 
Figs. 7 and 8 show the corresponding
$\Delta\sigma$ versus $\Delta y_{Zl}$ distributions.
The feature of the
narrow peak in the transverse mass distribution
for the 1 TeV heavy vector boson will make the signal stand
out. The resonant peak for $m_V=2.04$ TeV is much less distinct, and the
$\Delta y_{Zl}$ distribution provides auxiliary evidence of enhanced
longitudinal gauge boson production.

\section{Conclusions}

Direct measurements of $pp\rightarrow W^\pm Z^0\rightarrow l_1\nu_1
l_2\bar{l}_2$ at the LHC will constrain models with charged vector bosons
coupling to longitudinal $W$'s and $Z$'s. In the case of minimal technicolor
with one SU(2) doublet of technifermions, $N\geq 7$ in SU$(N)_{TC}$ for a 
significance equal to 5 or more given $10^{5}$ pb$^{-1}$ integrated
luminosity. If one considers the more general BESS model, an additional
parameter $a$ is introduced. The minimum $a$ ranges from 0.42--3.9 for
$M_V=1-2.04$ TeV if one requires a significance of 5. Rapidity correlations
demonstrate enhanced longitudinal gauge boson production and may be
an important tool in identifying the nature of nonstandard model
enhancements of weak gauge boson pair production.

Constraints on BESS and technicolor parameters from precision electroweak
measurements have been evaluated by Casalbuoni et al. \cite{constr}
and updated by Dominici in Ref. \cite{dominici}. 
Heavy vector bosons contribute to the oblique parameter $\epsilon_3$
through radiative corrections to the vector and axial vector
weak couplings.
Analyses of high energy experimental data \cite{altarelli} yield
$\epsilon_3^{exp}=(4.1\pm 1.4) \cdot 10^{-3}$. Standard model
radiative corrections, setting $m_H=1$ TeV, give 
$\epsilon_3^{th}=(6.6\pm 0.1)\cdot 10^{-3}$. In the case in which there is not
a fundamental Higgs boson, the radiative contribution to $\epsilon_3$
has the same form as the Higgs contribution, and is approximately equal
if one takes the cutoff at $\Lambda=1$ TeV \cite{constr}.
Taking the difference between the two values, nonstandard contributions
to $\epsilon_3$ which come in at the tree level, as is
the case here, are limited to
\begin{equation}
\epsilon_3<1.7\cdot 10^{-3}\ (4.5\cdot 10^{-3})
\end{equation}
at the $3\sigma$ (5$\sigma$) level. 

The standard technicolor model 
has contributions to $\epsilon_3$ from both heavy vector and heavy
axial vector bosons. The axial vector contributions come in with opposite
sign in a prescribed ratio. 
The net result is that
$M_{\rho_T}>2.38$ TeV ($3\sigma$) and $M_{\rho_T}>1.46$ TeV
(5$\sigma$), requiring the $N$ of $SU(N)$ to be $N<2.1$ or
$N<5.6$, respectively. 

In the case of the more general BESS model, there are no set relations between
vector and axial vector masses, so the limits are weaker.
In the case of degenerate heavy vector and axial vector bosons \cite{dominici},
$\epsilon_3\sim a M_W^4/M_V^4$. The limits in the degenerate case are not
especially relevant to the calculation here because we have not included
a degenerate axial vector boson contribution to the signal. In the case
where the heavy axial vector boson contribution to $\epsilon_3$ can be
ignored,
$a<0.27 (M_V/{\rm TeV})^2$ for the 3$\sigma$ case, and
$a<0.71 (M_V/{\rm TeV})^2$ for $5\sigma$ case.

The indirect limits on standard technicolor
and the BESS model in the absence of heavy axial vector 
boson corrections are in general stronger than
the potential LHC limits, but rely on the approximation of the 
radiative corrections. 
The LHC experiments will provide important direct constraints
on  the existence of charged heavy vector bosons.

\acknowledgements
Work supported in part
by National Science Foundation Grant No.
PHY-9507688 and PHY-9802403.

\begin{table}
\label{tab:sigma}
\caption{Cross section for $pp\rightarrow W^\pm Z^0\rightarrow
l_1\nu_1 l_2\bar{l}_2$ at the LHC for
$m_T\geq 0.5$ TeV, in units of $10^{-4}$ pb.
Rapidity cuts $|y_Z|,\, |y_{l_1}|<2.5$ were applied.}
\begin{tabular}{cc}
Polarization  & $\sigma$ \\
\tableline
\ TT  & 4.62    \\ 
\ TL  &  0.23   \\
\ LT  &  0.25    \\
\ LL  &  1.66   \\
\ LL-TC8 & 8.19  \\
\ LL-TC4 & 2.98\\
\ LL-TC3 & 2.35 \\ 
\end{tabular}
\end{table}

\begin{table}
\label{tab:lmin}
\caption{Minimum luminosity
for the detection of charged technirhos ($a=2$) at the LHC
for a calculated significance greater than 5, assuming one channel
for each boson decay.}

\begin{tabular}{ l  c }

$M_V$ [TeV] & ${\cal L}_{\rm min}$ [$10^5$ pb$^{-1}$] \\ \hline
1.25 $(N=8)$ & 0.6 \\ 
1.34 $(N=7)$ & 0.9 \\ 
1.44 $(N=6)$ & 1.5 \\ 
1.58 $(N=5)$ & 2.3 \\ 
1.76 $(N=4)$ & 5.9 \\ 
2.04 $(N=3)$ & 17.4 \\ 
\end{tabular}
\end{table}

\begin{table}
\label{tab:amin}
\caption{Limits on $a$  
for $pp\rightarrow W^\pm Z^0\rightarrow
l_1^\pm \nu_1 l_2^\pm l_2^\mp$ at the LHC with $10^5$
pb$^{-1}$ of luminosity. 
We have imposed
rapidity cuts $|y_Z|,\, |y_{l_1}|<2.5$. Transverse mass cuts
are fixed to maximize the significance of the signal.}

\begin{tabular}{ l  c }
$M_V$ [TeV] & $a_{\rm min}$ \\ \hline
1.00  & 0.42\\ 
1.25  & 1.4 \\ 
1.76  & 3.4 \\ 
2.04  & 3.9 \\ 
\end{tabular}
\end{table}

\begin{figure}
\vskip 0.5in
{\psfig{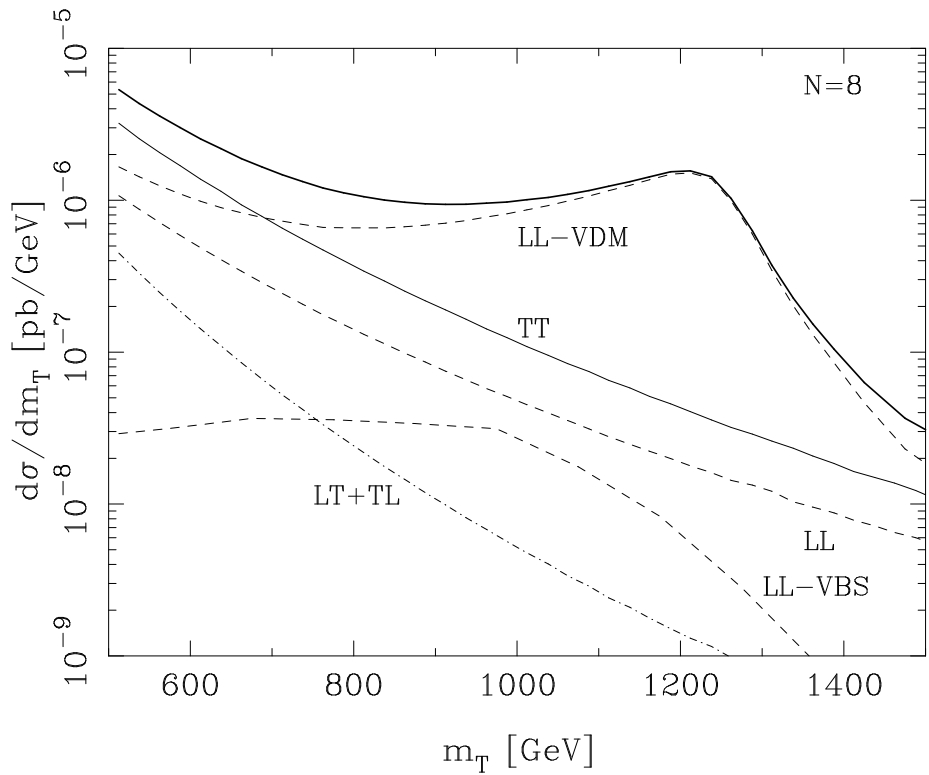}}
\vskip 0.3in
\caption{Differential cross section, including technirho 
(a=2) contributions,
as a function of $m_T$ for
$pp\rightarrow W^\pm Z^0\rightarrow l_1\nu_1 l_2^+ l_2^-$
at the LHC. The solid line is the sum of the transverse boson final
states (TT), the dashed line labeled LL is the standard model longitudinal
pair distribution and the mixed transverse-longitudinal pair (TL+LT)
is shown by the dot-dashed line. The technicolor contribution via
$q\bar{q}'$ annihilation is 
shown by the dashed line labeled by LL-VDM, while the vector boson scattering
contribution is labeled LL-VBS.
The technirho contribution assumes $N=8$ ($M_{\rho_T}=1.25$ TeV).
}
\end{figure}
\vskip 0.3in

\begin{figure}
\vskip 0.5in
\psfig{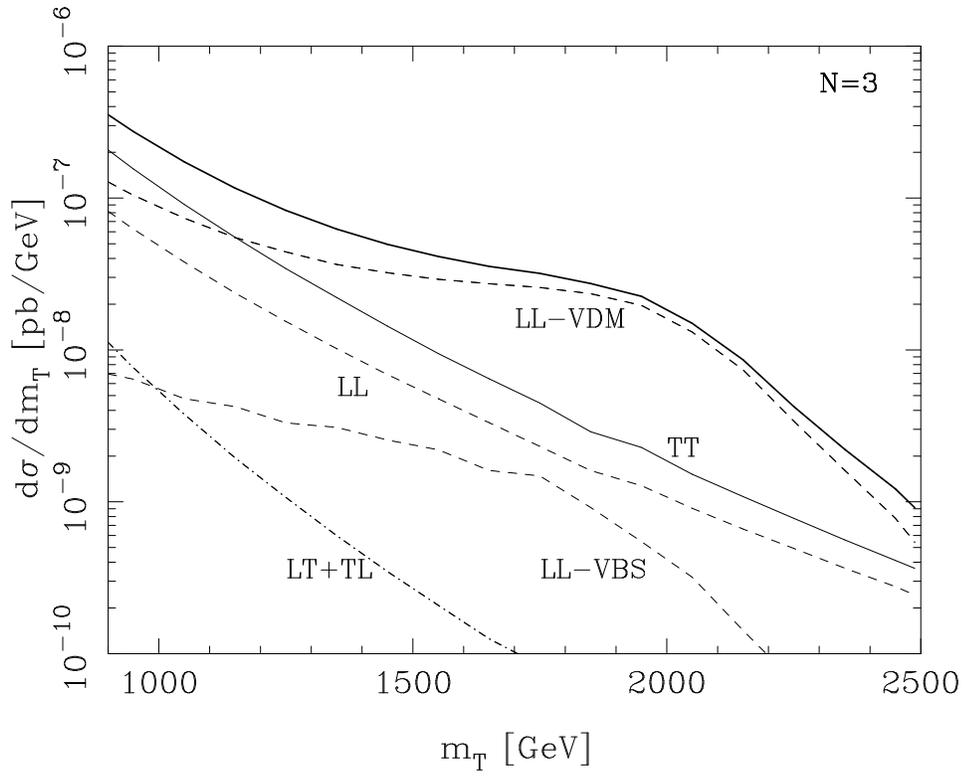}
\vskip 0.3in
\caption{As in Fig. 1, here with $N=3$ ($M_{\rho_T}=2.04$ TeV).}
\end{figure}
\vskip 0.3in

\begin{figure}
\vskip 0.5in
\centerline{\psfig{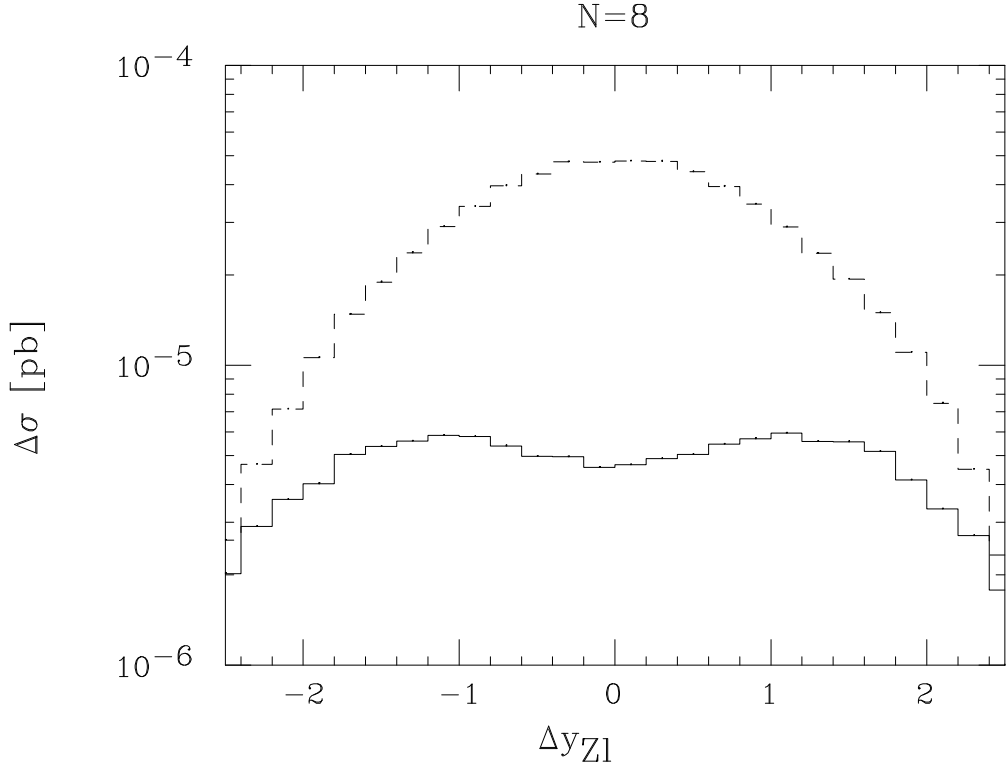}}
\vskip 0.3in
\caption{Cross section, per bin of 0.2 in rapidity, as a function  of
$\Delta y_{Zl}$ for $725
\ {\rm GeV}\leq m_T\leq 1600$ GeV. 
The solid histogram shows the standard
model result. The dashed histogram shows the signal plus background
for $N=8$ minimal technicolor.
}
\end{figure}
\vskip 0.3in

\begin{figure}
\vskip 0.5in
\centerline{\psfig{figure=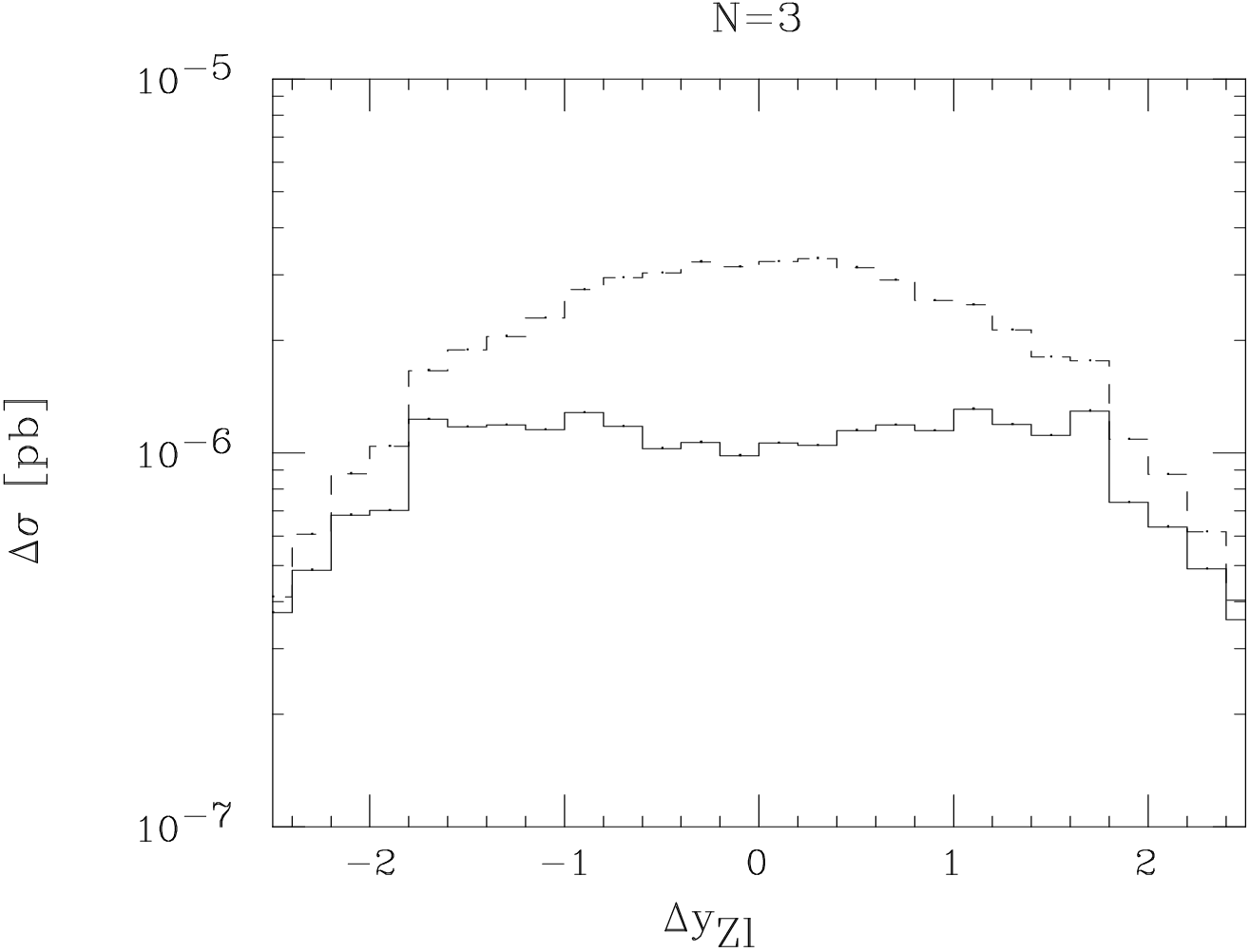,height=4.0in,angle=90}}
\vskip 0.3in
\caption{Cross section, per bin of 0.2 in rapidity, as a function  of
$\Delta y_{Zl}$ for $1075\ {\rm GeV}
\leq m_T\leq 2400$ GeV. The solid histogram shows the standard model
result. The dashed histogram shows the signal plus background for $N=3$
minimal technicolor.}
\end{figure}
\vskip 0.3in

\begin{figure}
\vskip 0.5in
\centerline{\psfig{figure=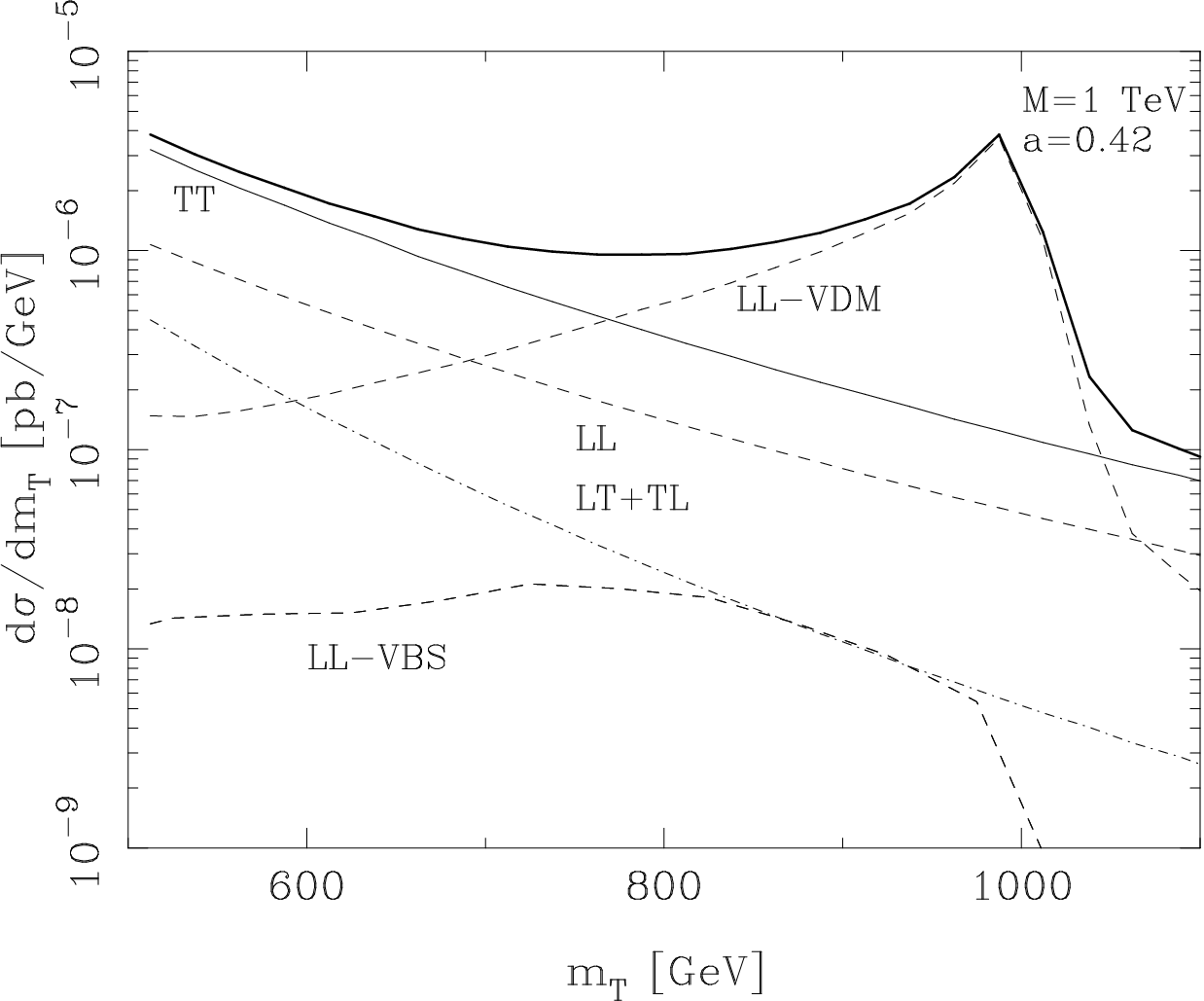,height=4.0in,angle=270}}
\vskip 0.3in
\caption{Differential cross section as a function of $m_T$ for
$M_V=1.00$ TeV and $a=0.42$. The curves are labeled as in Figure 1.
}
\end{figure}
\vskip 0.3in

\begin{figure}
\vskip 0.5in
\centerline{\psfig{figure=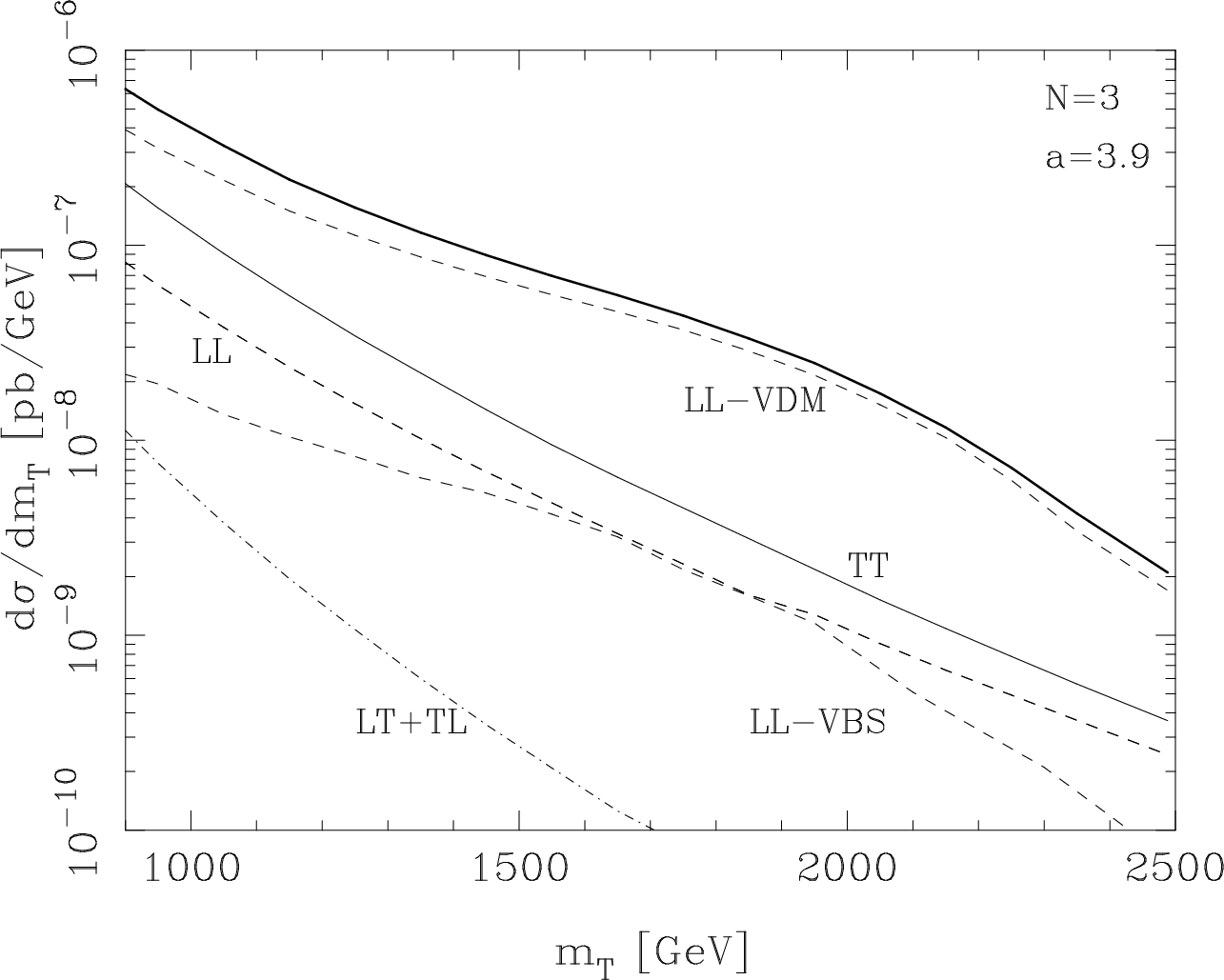,height=4.0in,angle=270}}
\vskip 0.3in
\caption{Differential cross section as a function of $m_T$ for
$M_V=2.04$ TeV and $a=3.9$. The curves are labeled as in Figure 1.
}
\end{figure}
\vskip 0.3in

\begin{figure}
\vskip 0.5in
\centerline{\psfig{figure=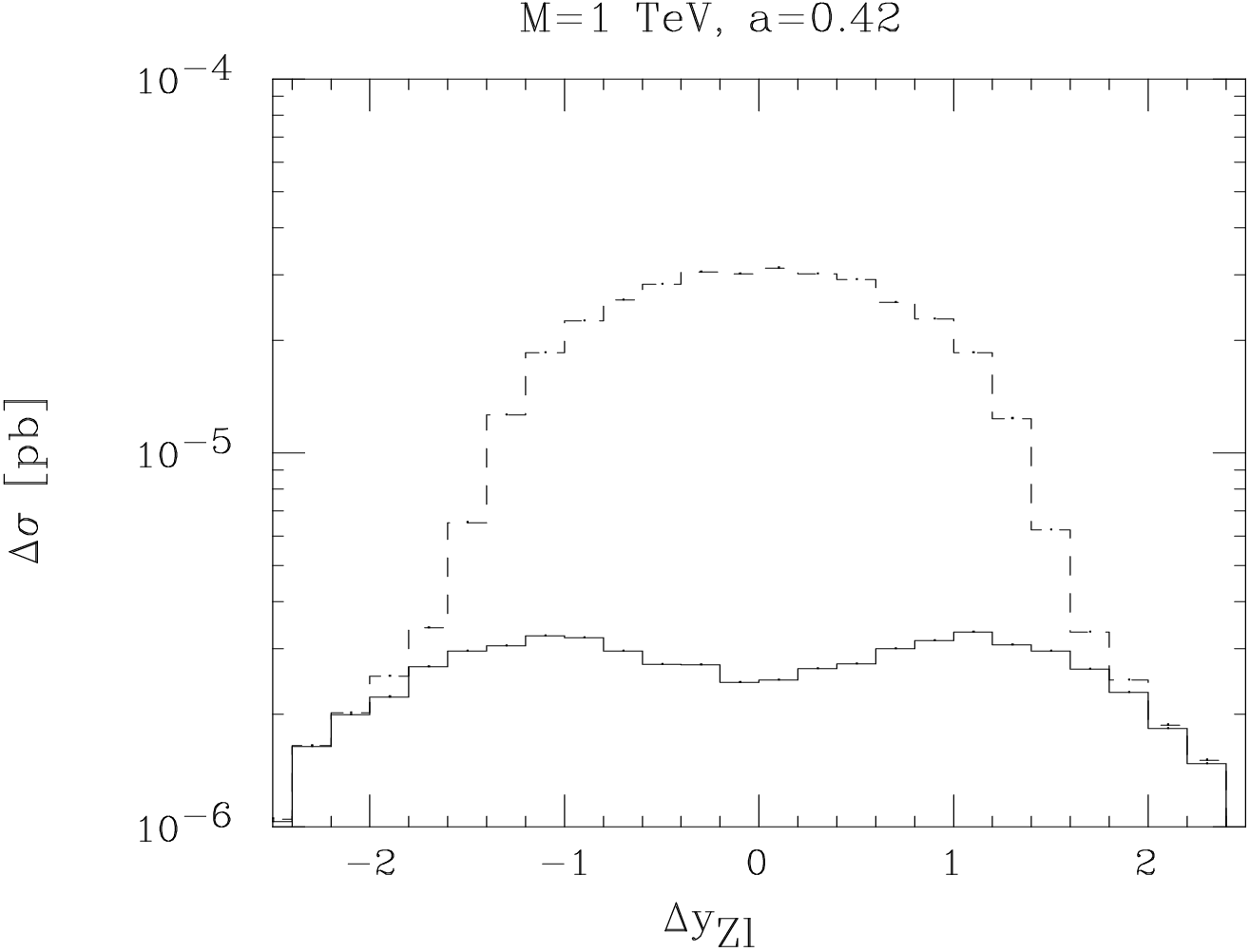,height=4.0in,angle=90}}
\vskip 0.3in
\caption{Cross section, per bin of 0.2 in rapidity, as a function  of
$\Delta y_{Zl}$ for $M=1.00$ TeV, $a=0.42$ and $800\ {\rm GeV}\leq m_T
\leq 1050$ GeV.
The solid histogram shows the standard
model result. The dashed histogram shows the signal plus background
for $N=8$ minimal technicolor.
}
\end{figure}
\vskip 0.3in

\begin{figure}
\vskip 0.5in
\centerline{\psfig{figure=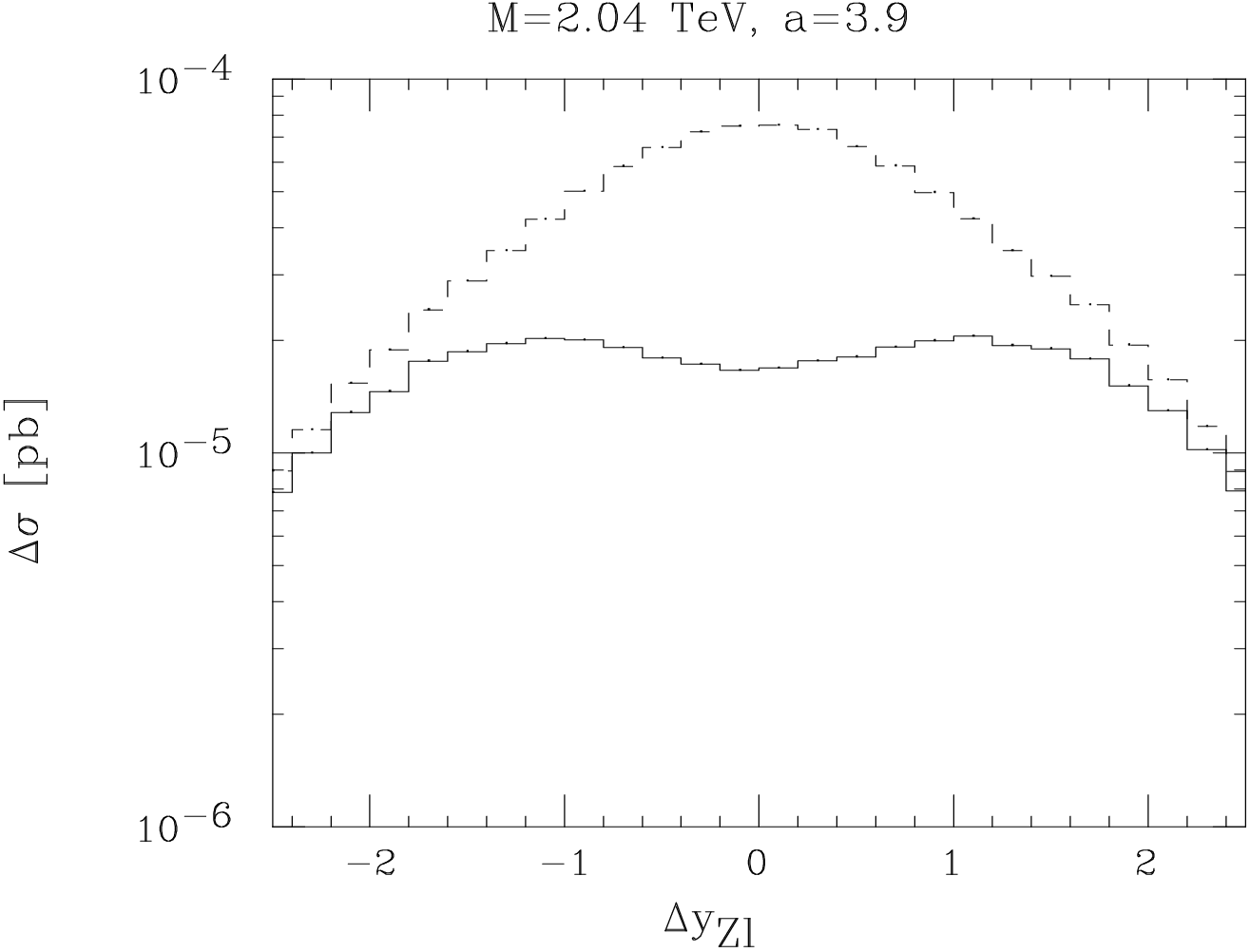,height=4.0in,angle=90}}
\vskip 0.3in
\caption{Cross section, per bin of 0.2 in rapidity, as a function  of
$\Delta y_{Zl}$ for $M=2.04$ TeV, $a=3.9$ and $500$ GeV$\leq m_T\leq 2500$
GeV.
The solid histogram shows the standard
model result. The dashed histogram shows the signal plus background
for $N=8$ minimal technicolor.
}
\end{figure}
\vskip 0.3in

\end{document}